\documentclass[12pt]{iopart}
\usepackage{iopams}  
\usepackage{graphicx}
\usepackage{cite}

\begin{document}

\title[Shortcut to adiabaticity in full-wave optics]{Shortcut to adiabaticity in full-wave optics for ultra-compact waveguide junctions}

\author{Giuseppe Della Valle$^{1,2}$, Gerardo Perozziello$^3$, Stefano Longhi$^{1,2}$}

\address{$^1$ Dipartimento di Fisica, Politecnico di Milano, Piazza L. da Vinci 32, I-20133 Milano, Italy}
\address{$^2$ Istituto di Fotonica e Nanotecnologie, CNR, Piazza L. da Vinci 32, I-20133 Milano, Italy}
\address{$^3$ Department of Experimental and Clinical Medicine, BioNEM lab, University {\it Magna Graecia} of Catanzaro, 88100 Catanzaro, Italy}
\ead{giuseppe.dellavalle@polimi.it}

\vspace{10pt}
\begin{indented}
\item[]May 2016
\end{indented}

\begin{abstract}
We extend the concept of shortcut to adiabaticity to full-wave optics and provide application to the design of an ultra-compact waveguide junction. In particular, we introduce a procedure allowing one to synthesize a purely dielectric optical potential that precisely compensates for the non-adiabatic losses of the transverse-electric fundamental mode in \textit{any} (sufficiently regular) two-dimensional waveguide junction. Our results are corroborated by finite-element method numerical simulations in a P\"oschl-Teller waveguide mode expander.
\end{abstract}

%
\vspace{2pc}
\noindent{\it Keywords}: Inhomogeneous optical media; Gradient-index media; Integrated optics devices.
%
%
\maketitle
%
%

The adiabatic theorem guarantees that if a system changes sufficiently slowly its eigenstates smoothly evolve without coupling to each others \cite{Born_ZPA_1928, Kato_JPSJ_1950, Griffiths_book}. This concept has been widely exploited in many different areas of physics, from quantum mechanics to optics \cite{Menchon_Review_2016}. As example, a gently-tapered waveguide allows one to efficiently convert a narrow waveguide mode into a broader one (or viceversa), which is a key functionality in optical interconnects \cite{Jedrzejewski_EL_1986, Love_EL_1986, Birks_JLT_1992}. Several photonic structures based on the adiabatic evolution of light waves have been reported \cite{Hope_OE_2013,Love_OQE_1996,Adar_JLT_1992,Yerolatsitis_OE_2014,Dreisow_APL_2009,Longhi_PRE_2006, Ciret_OL_2012, Menchon_LSA_2013,Mrejen_NC_2015}, including asymmetric y-branch couplers \cite{Love_OQE_1996}, 4-port filters and multiplexers \cite{Adar_JLT_1992, Yerolatsitis_OE_2014}, broadband splitters and filters \cite{Dreisow_APL_2009, Longhi_PRE_2006, Ciret_OL_2012, Menchon_LSA_2013}, and densely-packed sub wavelength waveguides \cite{Mrejen_NC_2015}, just to mention a few. Adiabatic protocols are very robust against variation of parameters \cite{Vitanov_submitted} but suffer from a fundamental limitation in terms of \textit{speed}, which in optics translates into a minimum length of the device, thus posing a major issue for its exploitation in integrated optics. 

A technique leading to a substantial speeding of the adiabatic evolution is referred to as Shortcut To Adiabaticity (STA). The concept of STA was originally introduced in quantum mechanics, and refers to a family of methods including, among others,  Counterdiabatic or Transitionless Tracking \cite{Demirplak_JCPA_2003, Demirplak_JCPB_2005, Demirplak_JCP_2008, Berry_JPA_2009, Bason_NP_2012}, Invariant-Based Inverse Engineering \cite{Chen_PRL_2010, Torrontegui_PRA_2011}, and Fast-Forward Approach \cite{Masuda_PRSA_2010, Masuda_PRA_2011, Torrontegui_PRA_2012} (see also Ref.~\citenum{Torrontegui_AAMOP_2013} and references therein). Very recently, the STA has been reported in other physical contexts, including non-hermitian quantum physics \cite{Torosov_PRA_2013,Torosov_PRA_2014} and, most interestingly, optics \cite{Lin_OE_2012, Stefanatos_PRA_2014, Tseng_OL_2014, Yeih_PTL_2014, Paul_PRA_2015}. The exploitation of STA in optics can led to much more compact waveguide devices as compared to the corresponding adiabatic structures, which is very beneficial to device integration. However, the optical STA has been limited to finite-dimensional systems where the dynamics of light transport is reduced to coupled-mode-equations (CMEs) formalism, allowing a precise mimicking of STA protocols already developed in quantum mechanics.

In this letter we extend the optical STA to full-wave problems for the Helmholtz equation, i.e.~to an infinite dimensional system. The analysis is pursued in a two dimensional space and for transverse electric (TE) fields. Our optical STA is exemplified by designing an efficient ultra-compact waveguide expander, where the radiation losses induced by shortening of an adiabatic configuration are compensated by superposing a suitable optical potential.

We consider an optical medium with translational invariance along one direction, $y$, and electromagnetic waves propagating in the $xz$ plane (a so called \textit{in-plane} wave propagation problem). For isotropic media the TE-TM decomposition holds, allowing one to reduce the general case to two decoupled scalar diffraction problems, one for the TE polarization (non zero electric field out of the $xz$ plane) and one for the TM polarization (non zero magnetic field out of the $xz$ plane). We limit our analysis to the TE case but, in principle, a similar approach can be pursued for the TM case as well (even though with a more complicated numerical approach that should deserve a dedicated study). In the time-harmonic regime for TE waves, the non zero electric field component $E(x,z)$ obeys the Helmholtz equation:
\begin{equation}
E_{xx}+E_{zz}+(\omega^2/c^2)\epsilon(x,z) E = 0,
\end{equation}
where $\omega$ is the optical pulsation, $c$ is the speed of light in vacuum, $\epsilon(x,z)$ is the relative permittivity of the medium and the subscript denotes partial derivative with respect to the space coordinates. 
Let's now consider the special case of a dielectric function having a very slow and smooth variation along $z$ (the optical axis) from $z = z_1$ to $z = z_2$, and being $z$-invariant out of the $[z_1,z_2]$ interval. The section between $z_1$ and $z_2$ is thus interpreted as a junction between an input ($z<z_1$) waveguide and an output ($z>z_2$) waveguide [Fig.~1(a)]. If the junction fulfills adiabatic conditions (see below), the electric field evolves adiabatically along the structure:
\begin{equation}
E(x,z) = A(x,z) e^{ i \int_{z_1}^z dz' n(z') \omega/c },
\end{equation}
where $A(x,z)$ and $n(z)$ are, respectively, the field profile and effective index of the \textit{local normal mode} at around the considered $z$ coordinate. More precisely, if we denote the dielectric function of the adiabatic structure as $\epsilon_A(x,z)$, $A$ and $n$ obey the following eigenvalue equation parametrized by $z$:
\begin{equation}
A_{xx}+(\omega^2/c^2)\epsilon_A(x,z) A = (\omega^2/c^2)n^2 A.
\end{equation}
If the optical potential $\epsilon_A(x,z)$ is purely real, the adiabatic waveguide junction is lossless, but ought to be very long in order to fulfill adiabatic conditions. The latter can be derived by inserting Eq. (2) into Eq. (1), leading, in the most general case, to an optical potential $\epsilon(x,z)$ made of the sum of two terms, the adiabatic potential $\epsilon_A(x,z)$ plus an extra potential $\epsilon_{NA}(x,z)$ that accounts for deviations from adiabaticity (if any):
\begin{equation}
\epsilon_{NA}(x,z) = -\frac{c^2}{\omega^2}\frac{A_{zz}(x,z)}{A(x,z)}-i \frac{c}{\omega}\left[\frac{d n(z)}{d z}+2 n(z)\frac{A_z(x,z)}{A(x,z)}\right].
\end{equation}
Note that $\epsilon \simeq \epsilon_{A}$, i.e.~$\epsilon_{NA}$ is negligible with respect to $\epsilon_{A}$, provided that  the evolution of $A$ and $n$ along $z$ is slow enough, and this requires the waveguide junction (i.e. the distance between $z_1$ and $z_2$) to be several orders of magnitude longer than the optical wavelength $\lambda = 2\pi c/\omega$. In case of non adiabatic evolution of $\epsilon_A$, the adiabatic field $E_A(x,z)$ of Eq.~(2) is no more an approximated solution to Eq. (1), and the waveguide junction becomes lossy: the input mode couples to other guided modes (if any), including back propagating modes in the input section, and to radiating modes. 

\begin{figure}[b!]
\centerline{\includegraphics[width=8.5cm]{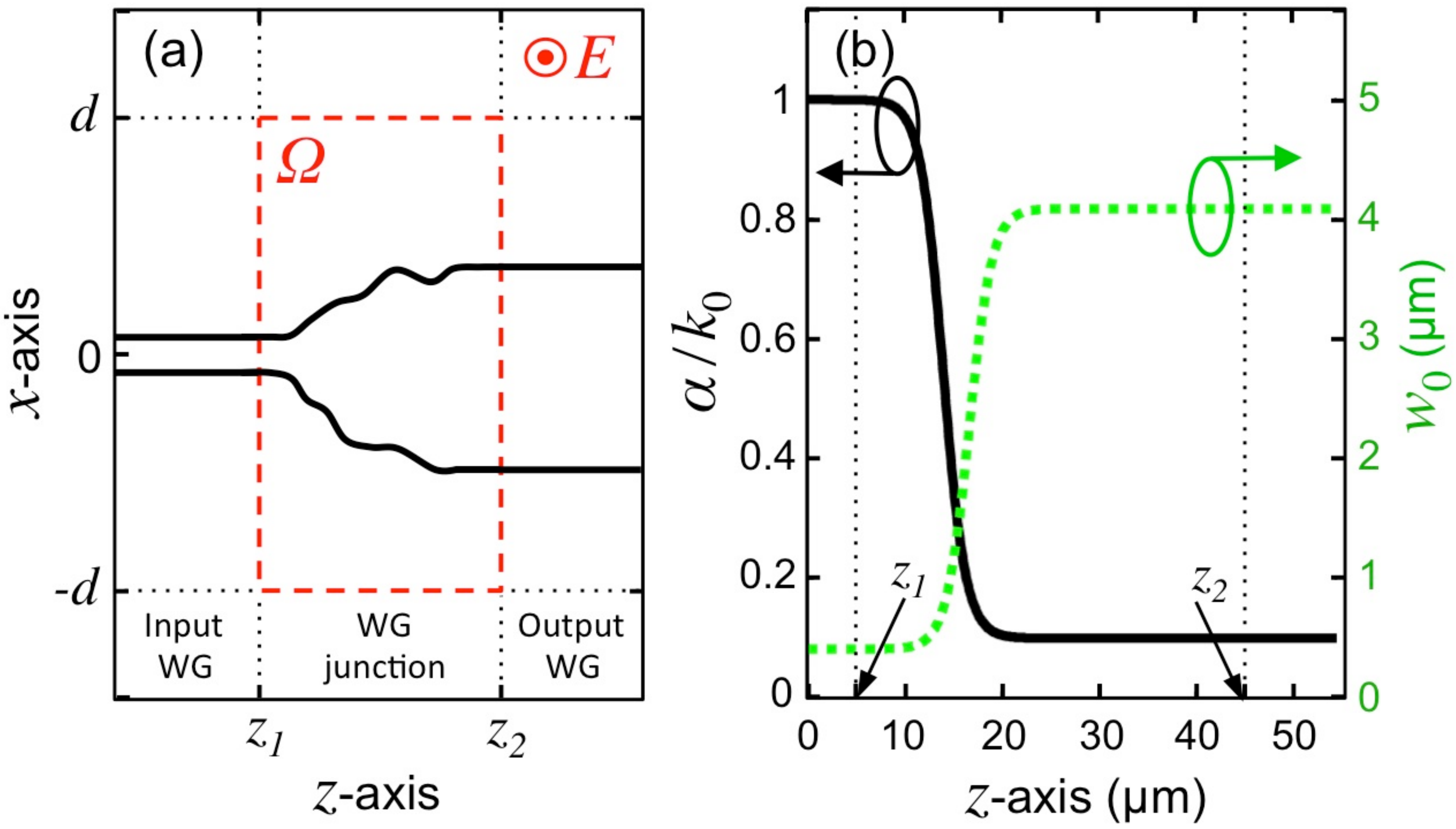}} 
\caption{(Color online) (a) Sketch of a generic waveguide junction for TE waves (non zero electric field component $E$ out of the $xz$ plane). (b) Evolution of the $\alpha$ parameter of Eq.~(11) (solid curve) in the P\"oschl-Teller waveguide structure defined by Eq.~(8) (see main text for parameters values). The dotted curve shows the evolution of the half width at $1/e$ of the local normal mode of the structure.}
\end{figure}

The seek for a shortcut to adiabaticity corresponds to the finding of a suitable optical potential $\Delta\epsilon_{STA}$ that superimposed to $\epsilon_{A}$ exactly compensates for the losses induced by non adiabaticity. Obviously, one can simply take $\Delta\epsilon_{STA} = -\epsilon_{NA}$. However, since $\epsilon_{NA}$ is complex valued and with both the real and the imaginary parts that depends on $x$ and $z$, this approach to STA is challenging to implement in a physical device, since it would require a complex control of spatially dispersed gain and loss in the medium. A viable approach to STA in the present context would provide a real valued $\Delta\epsilon_{STA}$. To this aim, we pursue an idea inspired to the streamlined version of the Fast-Forward STA introduced for the Schr\"odinger equation by Torrontegui \textit{et al.} \cite{Torrontegui_PRA_2012}. Let's assume, instead of Eq.~(2) a more general ansatz for the electric field propagating in the waveguide structure:
\begin{equation}
E(x,z) = A(x,z) e^{ i \int_{z_1}^z dz' n(z') \omega/c } \times e^{ i \phi(x,z)},
\end{equation}
\noindent with $\phi(x,z)$ a real-valued phase field to be determined under suitable boundary conditions. Note that above ansatz guarantees an evolution of the light intensity which is precisely that of a fast-forward adiabatic passage across the waveguide junction, since $|E(x,z)|^2 = |A(x,z)|^2$ as for the field of Eq.~(2). By inserting Eq.~(5) into Eq.~(1) and setting $\epsilon = \epsilon_A + \Delta\epsilon_{STA}$ a complex valued equation is retrieved for the complex valued STA potential $\Delta\epsilon_{STA}$. We can thus exploit the degree of freedom provided by the $\phi$ field to nullify the imaginary part of $\Delta\epsilon_{STA}$, which results into the following partial differential equation in the $\phi$ unknown:
\begin{equation}
A(x,z) \triangle \phi(x,z)+2\nabla A(x,z)\cdot \nabla \phi(x,z) = f(x,z),
\end{equation}
\noindent where $f(x,z) = -(\omega/c) [2 n(z) A_z(x,z) + A(x,z) d n(z)/ dz]$.
Provided that above equation can be solved with suitable boundary conditions, the resulting $\phi$ field is used to compute the non-zero part of $\epsilon_{STA}$, which is now purely real and reads as follows:
\begin{equation}
\Delta\epsilon_{STA}(x,z) = \frac{c^2}{\omega^2}\left\{[\nabla \phi(x,z)]^2-\frac{A_{zz}(x,z)}{A(x,z)}+2\frac{\omega}{c}n(z) \phi_z(x,z)\right\}.
\end{equation}
Interestingly, Eq.~(6) is a very-well known partial differential equation, that is the time-invariant convection-diffusion equation, for which several boundary value problems (BVP) have been extensively studied (see e.g. Ref.~\citenum{Salsa_book}). Note that, according to Eqs.~(6)-(7), the $\phi$ field is not uniquely defined, because the gauge transformation $\phi' = \phi + constant$ leads to the same STA potential. To guarantee the phase matching between the local normal mode and the propagating fields at the edges of the waveguide junction, one can impose $\phi_z(x,z_1) = \phi_z(x,z_2) = 0$. Also, one should require the optical STA potential to be localized within a certain distance $d$ from the optical axis. To do so we impose $\phi_x(d,z) = \phi_x(-d,z) = 0$. Above conditions identify a Neumann BVP for the $\phi$  field on the domain $\Omega = [z_1,z_2]\times[-d,d]$ in the $xz$ plane [Fig.~1(a)].
To provide gauge fixing and thus guarantee uniqueness of the solution, one of the Newmann boundary conditions was replaced with a Dirichlet one by setting $\phi(x,z_1) = 0$, which translates into a mixed BVP. It is worth pointing out that since in any waveguide structure the profile $A(x,z)$ of the fundamental (local) normal mode can be assumed to be positively defined, Eq.~(6) turns out to be elliptic in the wall $\Omega$ domain. Therefore, provided that the $f(x,z)$ field is smooth with its derivatives, standard methods of variational analysis based on the Lax-Milgram theorem guarantee that the mixed BVP above defined is well-posed \cite{Salsa_book}, and the search for a numerical solution can be pursued by, e.g., finite-element method (FEM) numerical analysis.

\begin{figure*}[t!]
\centerline{\includegraphics[width=15.5cm]{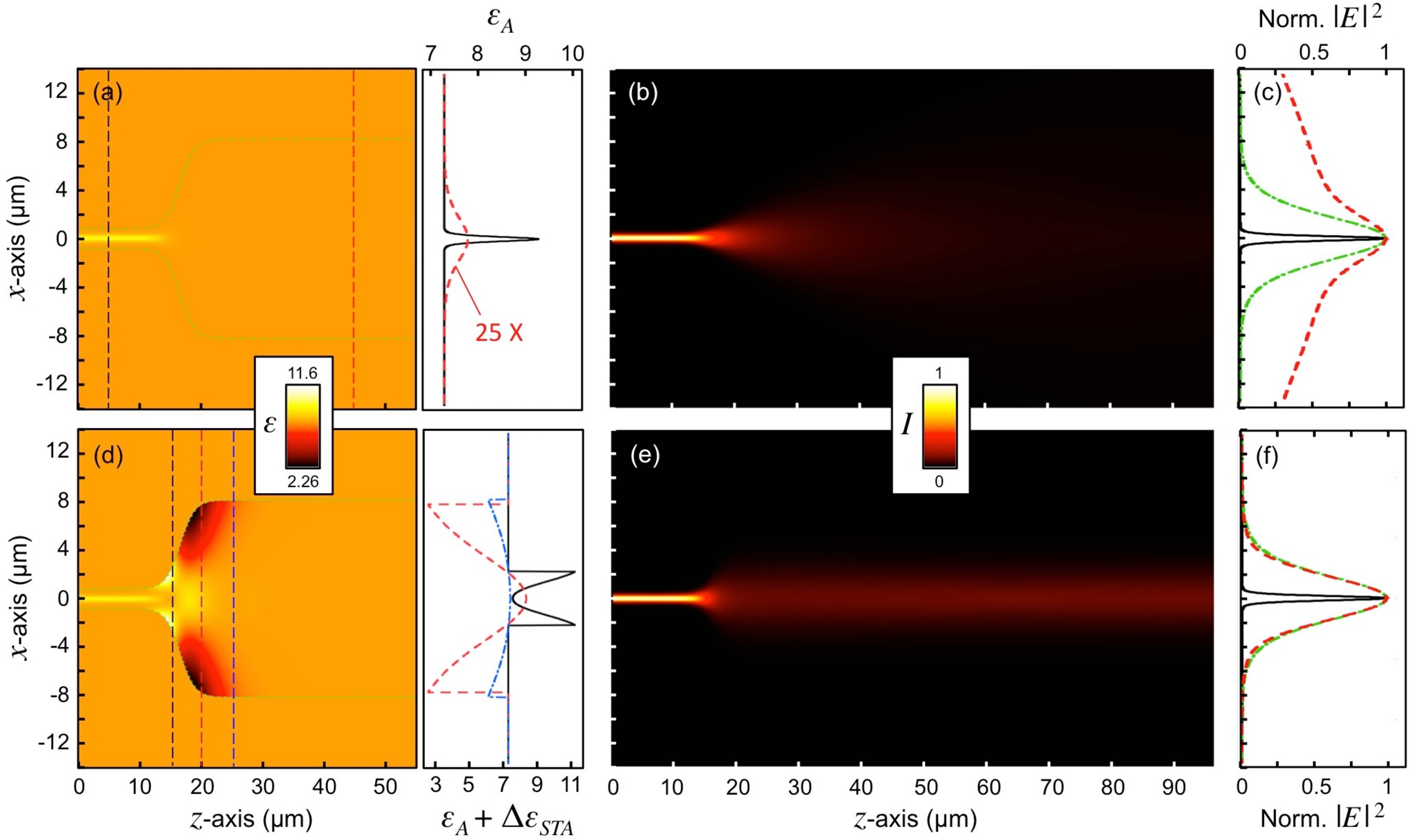}} 
\caption{(Color online) (a) Optical potential $\epsilon_A$ (pseudocolor map) of the adiabatic waveguide mode expander defined by Eqs.~(8)-(11). The waveguide junction (between the vertical dashed lines) is squeezed to only a 40 $\mu$m length (see main text for parameters values). Inset shows the cross-sections of $\epsilon_A$ at the input $z = z_1 = 5$~$\mu$m (solid curve) and at the output $z = z_2 = 45$~$\mu$m (dashed curve) of the waveguide junction (the latter being 25 times magnified for better reading). (b) Light intensity pattern (normalized to the peak) in the waveguide structure of panel (a) when the input waveguide mode is launched at $z = 0$. (c) Normalized electric field profile (modulus squared) at $z = 0$ (solid curve), and $z = 95$~$\mu$m (dashed curve). The profile of the output waveguide mode is also shown for comparison (dashdot line). (d), (e) and (f) same as (a), (b) and (c), respectively, for the waveguide mode expander with superimposed STA optical potential $\Delta\epsilon_{STA}$ designed to compensate deviations of $\epsilon_A$ to adiabaticity. The inset in panel (d) shows the cross-sections of $\epsilon_A+\epsilon_{STA}$ at $z = 15$~$\mu$m (solid curve), $z = 20$~$\mu$m (dashed curve) and $z = 25$~$\mu$m (dashdot curve). For the propagation simulations of panels (b)-(c) and (e)-(f) we employed Comsol Multiphysics 3.5 (RF module) with in-plane TE-waves analysis. Perfectly-matched layers were exploited to avoid back reflections at the boundaries of the integration domain.}
\end{figure*}

To exemplify our approach to the optical STA problem we apply above method to a P\"oschl-Teller (PT) waveguide and realize an ultra-compact mode-expander. A PT waveguide shows the following dielectric profile \cite{Lekner_AJP_2007}:
\begin{equation}
\epsilon_A (x,z) = n_s^2+2(c^2/\omega^2)\alpha^2(z){\rm sech}^2[\alpha(z) x]
\end{equation}
\noindent where $n_s$ is the refractive index of the waveguide cladding and $\alpha(z)$ is a $z$-dependent parameter exploited to control the evolution of the waveguide structure. As is well known \cite{Lekner_AJP_2007}, such a waveguide admits of a single mode whose adiabatic evolution according to Eq. (2) is ruled by:
\begin{eqnarray}
n(z) &=& \sqrt{n_s^2+\alpha^2(z)c^2/\omega^2},\\
A(x,z) &=& A_0\sqrt{n_1/\alpha_1} \sqrt{\alpha(z)/n(z)} \; {\rm sech}[\alpha(z) x].
\end{eqnarray}
In the above equations, $\alpha_1 = \alpha(z_1)$, $n_1 = n(z_1)$ (i.e.~the mode effective index of the input waveguide) and $A_0$ is an arbitrary constant (in V/m). 

To design a mode-expander for the PT waveguide one can evolve the $\alpha$ parameter from a high value $\alpha_1$ to a low value $\alpha_2$ along the $z$-axis. As example, we take:
\begin{equation}
\alpha(z) = \alpha_1 \left\{ \frac{r+1}{2}+\frac{r-1}{2}\frac{\tanh[(z-z_0)/\tau_0]}{\tanh(z_0/\tau_0)}\right\},
\end{equation}
where $r = \alpha_2/\alpha_1$, and $z_0$ and $\tau_0$ determine the spatial extent of the waveguide junction. An implementation of above formula with $\alpha_1 = \omega/c = 2\pi/\lambda$ (being $\lambda = 1.55$~$\mu$m the optical wavelength in vacuum), $r = 0.1$, $z_0 = 14$~$\mu$m and $\tau_0 = z_0/6$ is shown in Fig.~1(b) (solid line). This provides, by means of Eq.~(8) with $n_s = 2.7$, the optical potential of Fig.~2(a). To a high degree of accuracy such an optical potential turns out to be $z$-invariant for $z < z_1 = 5$~$\mu$m and for $z > z_2 = 45$~$\mu$m, the input and output waveguides respectively. The waveguide junction [between the vertical dashed lines in Fig.~2(a)] is thus only $40$~$\mu$m long (i.e.~$\sim25\lambda$). Also, note that the junction connects a very narrow waveguide with high permittivity core [solid curve in the inset of Fig.~2(a)] to a much broader waveguide with low permittivity core [dashed curve in the inset of Fig.~2(a)]. This structure largely violates the adiabatic conditions and the power launched on the mode of the input waveguide spreads on radiative modes after crossing the junction [Fig.~2(b)], with the result that the transmitted field at $z = 95$~$\mu$m [dashed curve in Fig.~2(c)] has poor superposition with the mode field of the output waveguide [dashdot curve in Fig.~2(c)]. 

A shortcut to adiabaticity for such a non adiabatic optical potential is realized by the dielectric function shown in Fig.~2(d), which is computed according to Eqs.~(6)-(7) and (9)-(11) by FEM numerical analysis with Comsol Multiphysics 3.5. To make the implementation of our method feasible in a high refractive index material like, e.g., silicon, the $\Delta\epsilon_{STA}$ retrieved from the $\phi$ computed by numerical integration of Eq.~(6) was truncated to zero out of the region where the adiabatic local normal mode of Eq.~(10) carries most of its energy. This is the region of the $xz$ plane  between the dotted curves in Fig.~2(a) and (d), where $|x|\leq 2w_0(z)$, being $w_0 = {\rm acosh}(e)/\alpha(z)$ (i.e.~the half width of the local normal mode at $1/e$), which is shown in Fig.~1(b) (dotted line). Note that the total optical potential $\epsilon_A + \Delta\epsilon_{STA}$ exhibits both positive and negative variations with respect to the cladding permittivity $n_s^2 = 7.29$ [inset in Fig.~2(d)]. To ascertain that the STA potential thus obtained is capable of compensating the radiation losses caused by deviations to adiabaticity of the $\epsilon_A$ potential, we performed FEM propagation analysis for the structure of Fig.~2(d) when launching the input waveguide mode. The resulting light intensity pattern along the structure is shown in Fig.~2(e). It well reproduces the adiabatic pattern detailed by Eqs.~(9)-(10), and the field transmitted by the junction [dashed line in Fig.~2(f)]  has very high (almost unitary) overlap with the mode field of the output waveguide [dashdot line in Fig.~2(f)]. Note that the junction is designed to operate at $\lambda = 1.55$~$\mu$m telecom wavelength and allows one to efficiently expand a narrow mode of $\sim0.8$~$\mu$m diameter to a 10 times broader one within a distance of $\sim 25\lambda$ [Fig.~1(b)].

Finally, it is worth noting that the gradient index structure of the waveguide junction of Fig.~2(d) can be fabricated by etching subwavelength holes in a silicon slab waveguide (see, e.g., Ref.~\citenum{Valentine_NM_2009} and references therein).

To conclude, we introduced in full-wave optics a method for shortcut to adiabaticity for an infinite dimensional system. Our method is developed for TE waves in a generic optical structure with a continuous translational invariance along one direction. The idea is applied to design an ultra-compact junction connecting a high permittivity core input waveguide to a weakly-guiding output waveguide of the kind of P\"oschl-Teller. The junction is designed by superposing to the squeezed adiabatic configuration a gradient index optical potential (the STA potential) that precisely compensates for the losses induced by the geometrical squeezing of the adiabatic structure. Further developments in the field should consider the case of TM polarization and eventually an extension of our method to three-dimensional optical structures.

\ack
This work was partially supported by the Fondazione Cariplo (Grant No. 2011-0338).

\end{document}